\documentclass[letterpaper,british,10pt,final,twocolumn]{IEEEtran}
\usepackage[T1]{fontenc}
\usepackage[latin1]{inputenc}
\usepackage{verbatim}
\usepackage{url}
\usepackage{amsmath}
\usepackage{graphicx}

\makeatletter

\providecommand{\tabularnewline}{\\}

\usepackage{epsfig}

\makeatother

\usepackage{babel}

\begin{document}

\title{A High-Quality Speech and Audio Codec With Less Than 10~ms Delay}

\author{Jean-Marc Valin, \textit{Member, IEEE}, Timothy B. Terriberry, Christopher
Montgomery, Gregory Maxwell\thanks{Jean-Marc Valin is with Octasic Inc., Montreal, Canada (part of the work was performed at the CSIRO ICT Centre, Marsfield, NSW, Australia). Christopher Montgomery is with RedHat Inc., USA. Gregory Maxwell is with Juniper Networks Inc., USA. All authors are with the Xiph.Org Foundation (email: jmvalin@ieee.org; tterribe@xiph.org; xiphmont@xiph.org; greg@xiph.org). The authors would like to thank all the volunteers who participated in the listening tests. \break
\copyright 2010 IEEE.  Personal use of this material is permitted. Permission from IEEE must be obtained for all other uses, in any current or future media, including reprinting/republishing this material for advertising or promotional purposes, creating new collective works, for resale or redistribution to servers or lists, or reuse of any copyrighted component of this work in other works.
}}
\maketitle
\begin{abstract}
With increasing quality requirements for multimedia communications,
audio codecs must maintain both high quality and low delay. Typically,
audio codecs offer either low delay or high quality, but rarely both.
We propose a codec that simultaneously addresses both these requirements,
with a delay of only 8.7~ms at 44.1~kHz. It uses gain-shape algebraic
vector quantisation in the frequency domain with time-domain pitch
prediction. We demonstrate that the proposed codec operating at 48~kbit/s
and 64~kbit/s out-performs both G.722.1C and MP3 and has quality
comparable to AAC-LD, despite having less than one fourth of the algorithmic
delay of these codecs.
\end{abstract}
\begin{center}\textbf{EDICS Category: SPE-CODI, AUD-ACOD}\end{center}


\section{Introduction\label{sec:Introduction}}

Meeting increasing expectations for video-conferencing and other communication
applications requires a high-quality, very low delay speech and audio
codec. Decreasing the delay both reduces the perception of acoustic
echo and enables new applications, such as remote music performances~\cite{Carot2006}.
Popular speech codecs such as AMR-WB~\cite{Bessette2002}, G.729.1~\cite{G.729.1},
and Speex~\cite{ValinAES2006} have a low-to-medium quality range,
do not support sampling rates above 16~kHz, and have total algorithmic
delays ranging from 15~ms to 30~ms. On the other hand, commonly
used audio codecs, such as MP3 and Vorbis~\cite{VorbisSpec}, can
achieve high quality but have delays exceeding 100~ms. None of these
codecs provide both high quality and very low delay. 

Since Code-Excited Linear Prediction (CELP)~\cite{Schroeder1985}
was proposed in the 1980s, it has been the most popular class of speech
coding algorithms. It is, however, generally limited to sampling rates
below 16 kHz. In the authors' experience and as reported in~\cite{Bessette2002},
CELP's noise shaping is difficult to control when the spectrum has
high dynamic range, as is common for sampling rates of 16~kHz and
above. To the authors' knowledge, CELP has not been applied to speech
codecs beyond a 16~kHz sampling rate. Even at 16~kHz, many CELP-based
codecs do not use CELP for the entire audio band. AMR-WB applies CELP
on a down-sampled 12.8 kHz signal, while G.729.1 uses an MDCT for
frequencies above 4~kHz~\cite{G.729.1}, and Speex encodes 16~kHz
speech using a Quadrature Mirror Filter and two CELP encoders~\cite{ValinAES2006}.

Unlike speech codecs, most audio codecs designed for music now use
the Modified Discrete Cosine Transform (MDCT). The algorithmic delay
of an MDCT-based codec is equal to the length of the window it uses.
Unfortunately, the very short time window required to achieve delays
below 10~ms does not give the MDCT sufficient frequency resolution
to model pitch harmonics. Most codecs based on the MDCT use windows
of 50~ms or more, although there are exceptions, such as G.722.1C
and AAC-LD, which use shorter windows.

We propose a new algorithm called the Constrained-Energy Lapped Transform
(CELT), detailed in Sections~\ref{sec:Code-Excited-Fourier-Transform}
and~\ref{sec:Quantisation}, that uses the MDCT with very short windows.
It explicitly encodes the energy of each spectral band, constraining
the output to match the spectral envelope of the input, thus preserving
its general perceptual qualities. It incorporates a time-domain pitch
predictor using the past of the synthesis signal to model the closely-spaced
harmonics of speech, giving both low delay and high resolution for
harmonic signals, just like speech codecs~\cite{Schroeder1985}.
For non-harmonic signals, the energy constraints prevent the predictor
from distorting the envelope of the signal, and it acts as another
vector quantisation codebook that only uses a few bits. The codec
has the following characteristics:
\begin{itemize}
\item a 44.1~kHz sampling rate,
\item a 8.7~ms algorithmic delay (5.8 ms frame size with 2.9 ms look-ahead),
\item high quality speech around 48~kbit/s, and
\item good quality music around 64~kbit/s.
\end{itemize}
We give the results of a number of experiments in Section~\ref{sec:Experiments-and-Results}.
We performed subjective listening tests against several other codecs,
and found CELT to equal or out-perform both G.722.1C and AAC-LD, while
achieving significantly less delay than all of them. We also performed
an objective analysis of the effect of transmission errors and found
CELT to be robust to random packet loss rates up to 5\% and bit error
rates (BER) as high as $3\times10^{-4}$ (0.03\%).

\section{Constrained-Energy Lapped Transform}

\label{sec:Code-Excited-Fourier-Transform}

One of the key issues with MDCT-based codecs is the time-frequency
resolution. For example, a codec proposed in~\cite{Norden2005} uses
a 35~ms window (17.5~ms frames) to achieve sufficient frequency
resolution to resolve the fine structure of pitch harmonics in speech.
CELT's very low delay constraint implies that it must use a very short
MDCT and hence has poor frequency resolution. To mitigate the problem,
we use a long-term predictor that extends far enough in the past to
model an entire pitch period. 

Another issue with using very short frames is that only a very small
number of bits is available for each frame. CELT must limit or eliminate
meta-information, such as that signalling bit allocation, and will
usually have just a few bits available for some frequency bands. For
that reason, we separate the coding of the spectral envelope from
the coding of the details of the spectrum. This ensures that the energy
in each frequency band is always preserved, even if the details of
the spectrum are lost.

CELT is inspired by CELP~\cite{Schroeder1985}, using the idea of
a spectrally flat ``excitation'' that is the sum of an adaptive (pitch)
codebook and a fixed (innovation) codebook. The excitation represents
the details of the spectrum after the spectral envelope has been removed.
However, unlike CELP, CELT mainly operates in the frequency domain
using the Modified Discrete Cosine Transform (MDCT), so the excitation
in CELT is a frequency-domain version of the excitation in CELP. Similarly,
the adaptive codebook is based on a time offset into the past with
an associated set of gains, and the innovation is the part of the
excitation that is not predicted by the adaptive codebook.

The main principles of the CELT algorithm are that
\begin{itemize}
\item the MDCT output is split in bands approximating the critical bands;
\item the encoder explicitly codes the energy in each band (spectral envelope)
and the decoder ensures the energy of the output matches the coded
energy exactly;
\item the normalised spectrum in each band, which we call the excitation,
is constrained to have unit norm throughout the process; and
\item the long-term (pitch) predictor is encoded as a time offset, but with
a pitch gain encoded in the frequency domain.
\end{itemize}
\begin{figure*}[t]
\begin{center}\includegraphics[width=0.8\textwidth]{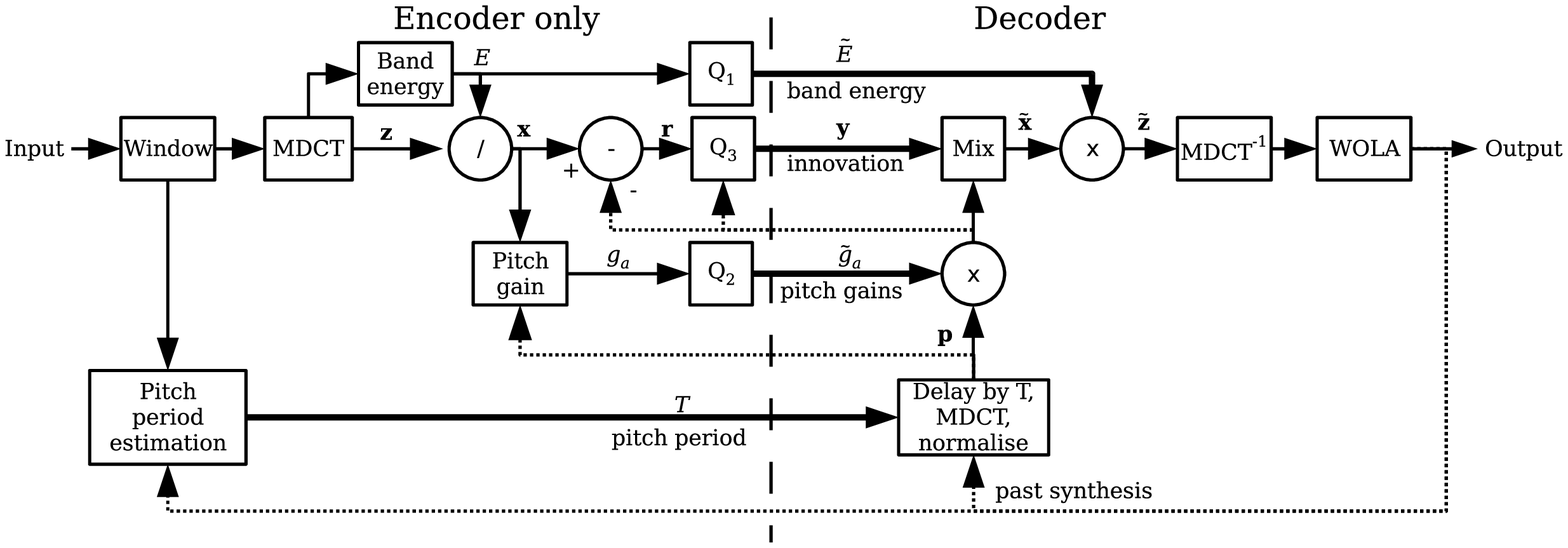}\end{center}

\caption{Overview of the CELT algorithm. The complete encoder includes the
decoder part because the encoding process refers to previously decoded
portions of the synthesis signal. Parameters transmitted to the decoder
are shown in bold and the parameters that are synchronised between
the encoder and the decoder are shown with dotted lines. Quantisers
are denoted by the $Q_{x}$ operators.\label{fig:Overview-of-CEFT}}

\end{figure*}

A block diagram of the CELT algorithm is shown in Fig.~\ref{fig:Overview-of-CEFT}.
The bit-stream is composed of 4 sets of parameters: the energy in
each band, the pitch period, the pitch gains, and the innovation codewords.
The most important variables are defined in Fig.~\ref{fig:Summary-of-variable}.

\begin{figure}
\begin{description}
\item [{$\alpha$}] inter-frame energy prediction coefficient
\item [{$\beta$}] inter-band energy prediction coefficient 
\item [{$\mu$}] mean energy in a band (fixed, computed offline)
\item [{$b$}] band index, loosely following the critical bands
\item [{$E$}] energy in band $b$ for frame $\ell$ (alternatively $E_{dB}$
in dB scale)
\item [{$\tilde{E}$}] quantised energy (alternatively $\tilde{E}_{dB}$
in dB scale)
\item [{$g_{a}$}] adaptive codebook gain
\item [{$\tilde{g}_{a}$}] quantised adaptive codebook gain
\item [{$g_{f}$}] fixed (innovation) codebook gain
\item [{$J$}] cost function for the innovation search
\item [{$K$}] number of pulses assigned to a band
\item [{$\ell$}] frame index
\item [{$L$}] length of the overlap (where the window is neither one nor
zero)
\item [{$N$}] number of MDCT samples in a band
\item [{$n_{k}$}] position of the $n^{th}$ innovation pulse
\item [{$\mathbf{p}$}] normalised adaptive code vector (pitch or folding)
\item [{$\mathbf{r}$}] residual after prediction (unquantised innovation)
\item [{$S$}] coarse energy quantiser resolution (6~dB)
\item [{\textmd{$s_{k}$}}] sign of the $k^{th}$ innovation pulse
\item [{$T$}] pitch period: time offset used for the long-term predictor
\item [{$V$}] number of pulse combinations
\item [{$w$}] window function
\item [{$\mathbf{x}$}] excitation: MDCT coefficients after normalisation
\item [{$\tilde{\mathbf{x}}$}] quantised excitation
\item [{$\mathbf{y}$}] quantised innovation
\item [{$\mathbf{z}$}] MDCT coefficients
\item [{$\tilde{\mathbf{z}}$}] quantised MDCT coefficients
\end{description}
\caption{Summary of variable definitions. Many of these variables have indices
$b$ and $\ell$, which are often omitted for clarity.\label{fig:Summary-of-variable}}

\end{figure}

\begin{figure}
\begin{center}\includegraphics[width=1\columnwidth]{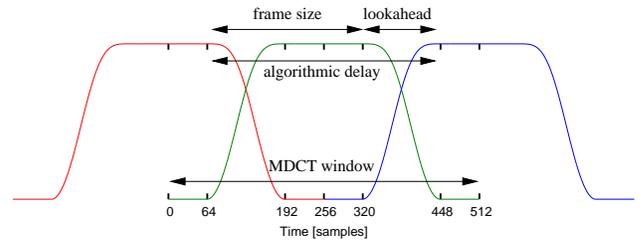}\end{center}\caption{Power-complementary windows with reduced overlap. The frame size is
256~samples, with~128 samples overlap. The total algorithmic delay
is 384 samples.\label{fig:Power-complementary-windows}}

\end{figure}

The signal is divided into 256 sample frames, with each MDCT window
composed of two frames. To reduce the delay, the overlap is only 128~samples,
with a 128-sample constant region in the centre and 64~zeros on each
side, as shown in Fig.~\ref{fig:Power-complementary-windows}. For
the overlap region, we use the Vorbis~\cite{VorbisSpec} codec's
power-complementary window:\begin{equation}
w\left(n\right)=\sin\left[\frac{\pi}{2}\sin^{2}\left(\frac{\pi\left(n+\frac{1}{2}\right)}{2L}\right)\right]\ ,\label{eq:vorbis-window}\end{equation}
where $L=128$ is the amount of overlap. Although a critically sampled
MDCT requires a window that is twice the frame size we reduce the
{}``effective overlap'' with the zeros on each side and still achieve
perfect reconstruction. This reduces the total algorithmic delay with
very little cost in quality or bit-rate. We use the same window for
the analysis process and the weighted overlap-and-add (WOLA) synthesis
process. %

\subsection{Bands and Energy}

CELT exploits the fact that the ear is mainly sensitive to the amount
of energy in each critical band. The MDCT spectrum is thus divided
into 20 bands of roughly one critical band each, although the lower
frequency bands are wider due to the low MDCT resolution. We refer
to these bands as the \emph{energy bands}. We normalise the MDCT spectrum
in each band and transmit the energy separately. Let $\mathbf{z}_{b}\left(\ell\right)$
be the MDCT spectrum in band $b$ at time frame $\ell$. Then the
normalised excitation in band $b$ is \begin{equation}
\mathbf{x}_{b}\left(\ell\right)=\frac{\mathbf{z}_{b}\left(\ell\right)}{\sqrt{E\left(b,\ell\right)}}\ ,\end{equation}
where $E\left(b,\ell\right)=\mathbf{z}_{b}^{T}\left(\ell\right)\mathbf{z}_{b}\left(\ell\right)$
is the energy in band $b$, so that $\mathbf{x}_{b}^{T}\left(\ell\right)\mathbf{x}_{b}\left(\ell\right)=1$.

A quantised version of the energy and the spectrum in each band is
transmitted to the decoder so that the signal can be recovered using\begin{equation}
\tilde{\mathbf{z}}_{b}=\sqrt{\tilde{E}\left(b,\ell\right)}\tilde{\mathbf{x}}_{b}\left(\ell\right)\ ,\end{equation}
where the quantised excitation $\tilde{\mathbf{x}}_{b}\left(\ell\right)$
still obeys $\tilde{\mathbf{x}}_{b}^{T}\left(\ell\right)\tilde{\mathbf{x}}_{b}\left(\ell\right)=1$.
This gain-shape approach has the advantage of preserving the spectral
envelope regardless of the bit-rate used to encode the {}``details''
of the spectrum. It also means that the spectral envelope $E\left(b,\ell\right)$
must be encoded at sufficient resolution, since $\tilde{\mathbf{x}}_{b}\left(\ell\right)$
cannot compensate for the quantisation error in $\tilde{E}\left(b,\ell\right)$.
This is unlike CELP codecs, where increasing the bit-rate of the excitation
can partially compensate for quantisation error in the LP coefficients.

From here on, unless we are processing multiple bands or multiple
frames at once, the frequency band $b$ and time frame $\ell$ are
omitted for clarity.

\subsection{Pitch Prediction}

\label{sub:Pitch-Prediction}

CELT uses pitch prediction to model the closely-spaced harmonics of
speech, solo instruments, or other highly periodic signals. By itself,
our short block transform is only capable of resolving harmonics if
the period is an exact multiple of the frame size. For any other period
length, the current window will contain a portion of the period offset
by some phase. We search the recently decoded signal data for a window
that covers the same portion of the period with the same phase offset.
While the harmonics will still not resolve into distinct MDCT bins,
for periodic inputs the predictor will produce the same pattern of
energy spreading.

The pitch predictor is specified by a period defined in the time domain
and a set of gains defined in the frequency domain. The pitch period
is the time offset to the window in the recent synthesis signal history
that best matches the current encoding window. We estimate the period
using the frequency-domain generalised cross-correlation between the
zero-padded input window and the last $L_{p}=1024$ decoded samples~\cite{Knapp1976}.
We use a weight function to normalise the response at each frequency
by the magnitude of the input window's spectrum, which is a crude
substitute for the perceptual weight CELP uses when computing time-domain
cross-correlation. Because the delayed signal used for pitch cannot
overlap with the current frame, the minimum delay possible is $N+L$
(384 samples). This corresponds to a 115 Hz fundamental, meaning that
for female speakers the estimated period is usually a multiple of
the real pitch period. Since the maximum period is equal to $L_{p}$,
there are $L_{p}-N-L+1=641$ possible time offsets.

Given the period, we compute the MDCT on the windowed, delayed synthesis
signal and normalise it to have unit magnitude in each band. We apply
the gain to the normalised signal, $\mathbf{p}$, in the frequency
domain, allowing us to vary the gain as a function of frequency. We
compute the gain in each band between 0 and 8~kHz as\begin{equation}
g_{a}=g_{damp}\frac{\mathbf{x}^{T}\mathbf{p}}{\mathbf{p}^{T}\mathbf{p}}=g_{damp}\mathbf{x}^{T}\mathbf{p}\ ,\end{equation}
where $g_{damp}$ is the gain \emph{damping factor} that also acts
as an upper bound for the gain, since $\mathbf{x}^{T}\mathbf{p}\leq1$.
Above 8~kHz, the adaptive codebook uses spectral folding from the
current frame, as described in Section \ref{sub:Sparseness-prevention}.
Because both $\mathbf{x}$ and $\mathbf{p}$ are normalised, the optimal
gain may never exceed unity. This is unlike the CELP algorithm, where
the optimal pitch gain may be greater than unity during onsets, as
mentioned by~\cite{Bessette2002}, which limits the gain to 1.2 to
prevent unstable behaviour. We limit the gain to $g_{damp}=0.9$ to
improve robustness to packet loss, not to avoid instability. 

We apply the pitch gain in the frequency domain to account for the
weakening of the pitch harmonics as the frequency increases. While
a 3-tap time-domain pitch gain~\cite{Chen1995} works with an 8~kHz
signal, it does not allow sufficient control of the pitch gain as
a function of frequency for a 44.1~kHz signal.

\subsection{Innovation}

In a manner similar to the CELP algorithm, the adaptive codebook and
fixed codebook contributions for a certain frequency band are combined
with\begin{equation}
\widetilde{\mathbf{x}}=\tilde{g}_{a}\mathbf{p}+g_{f}\mathbf{y}\ ,\label{eq:mixing}\end{equation}
where $\tilde{g}_{a}$ is the quantised gain of the adaptive codebook
contribution $\mathbf{p}$ and $g_{f}$ is the gain for the fixed
codebook contribution $\mathbf{y}$. Unlike CELP, the fixed codebook
gain $g_{f}$ does not need to be transmitted. Because of the constraint
$\tilde{\mathbf{x}}_{b}^{T}\tilde{\mathbf{x}}_{b}=1$ and knowing
that $\mathbf{p}^{T}\mathbf{p}=1$, the fixed codebook gain can be
computed as\begin{equation}
g_{f}=\frac{\sqrt{\tilde{g}_{a}^{2}\left(\mathbf{y}^{T}\mathbf{p}\right)^{2}+\mathbf{y}^{T}\mathbf{y}\left(1-\tilde{g}_{a}^{2}\right)}-g_{a}\mathbf{y}^{T}\mathbf{p}}{\mathbf{y}^{T}\mathbf{y}}\ .\label{eq:fixed-codebook-gain}\end{equation}
If $\tilde{g}_{a}=0$, then~\eqref{eq:fixed-codebook-gain} simplifies
to $g_{f}=1/\sqrt{\mathbf{y}^{T}\mathbf{y}}$, which only ensures
that $g_{f}\mathbf{y}$ has unit norm.

\section{Quantisation}

\label{sec:Quantisation}

This section describes each of the quantisers used in CELT. As shown
in Fig.~\ref{fig:Overview-of-CEFT}, we use three different quantisers:
one for the band energies, one for the pitch gains, and one for the
innovation. We entropy code the quantised results with a range coder~\cite{Mart79},
a type of arithmetic coder that outputs eight bits at a time. For
other quantisers, entropy coding is not necessary, but we still use
the range coder because it allows us to allocate a fractional number
of bits to integers whose size is not a power of two. For example,
using a range coder we can encode an integer parameter ranging from
0 to 2 using three symbols of probability $1/3$. This requires only
$\log_{2}3\approx1.59$~bits instead of the 2~bits necessary to
encode the integer directly.

\subsection{Band Energy Quantisation ($Q_{1}$)}

Efficiently encoding the energies $E\left(\ell,b\right)$ requires
eliminating redundancy in both time and frequency domain. Let $E_{dB}\left(\ell,b\right)$
be the log-energy in band $b$ at time frame $\ell$ We quantise this
energy as\begin{equation}
q_{b}\left(\ell\right)=\left\langle \frac{E_{dB}\left(\ell,b\right)-\mu_{b}-\alpha\tilde{E}_{dB}\left(\ell-1,b\right)-D\left(\ell,b\right)}{S}\right\rangle \ ,\label{eq:Equant-prediction}\end{equation}
\begin{align}
\tilde{E}_{dB}\left(\ell,b\right) & =S\left(q_{b}\left(\ell\right)+\mu_{b}+\alpha\tilde{E}_{dB}\left(\ell-1,b\right)+D\left(\ell,b\right)\right)\ ,\\
D\left(\ell+1,b\right) & =D\left(\ell,b\right)+\mu_{b}+\left(1-\beta\right)Sq_{b}\left(\ell\right)\ ,\end{align}
where $\left\langle \cdot\right\rangle $ denotes rounding to the
nearest integer, $q_{b}\left(\ell\right)$ is the encoded symbol,
$\mu_{b}$ is the mean energy for band $b$ (computed offline), $S$
is the quantisation resolution in dB, $\alpha$ controls the prediction
across frames, and $\beta$ controls the prediction across bands.
Not taking into account the fact that the prediction in~\eqref{eq:Equant-prediction}
is based on the quantised energy, the 2-D \emph{z}-transform of the
prediction filter is\begin{equation}
A\left(z_{\ell},z_{b}\right)=\left(1-\alpha z_{\ell}^{-1}\right)\cdot\frac{1-z_{b}^{-1}}{1-\beta z_{b}^{-1}}\ .\end{equation}

\begin{figure}
\begin{center}\includegraphics[width=0.8\columnwidth]{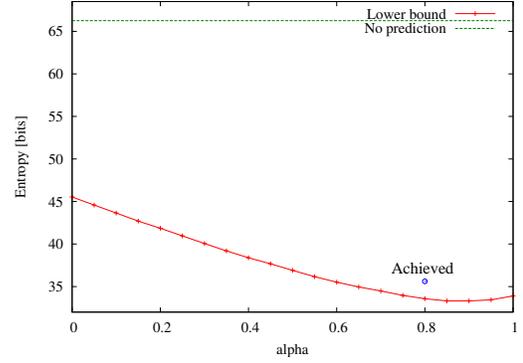}\end{center}\caption{Entropy of the energy prediction error (quantised with 6~dB resolution)
as a function of the inter-frame prediction coefficient $\alpha$
using the corresponding optimal value of $\beta$. The \emph{lower
bound} curve is a measurement of the entropy based on the probabilities
measured on the same data.\label{fig:Entropy1}}

\end{figure}

To find the optimal values for $\alpha$ and $\beta$, we measure
the entropy in the prediction error prior to encoding. Fig.~\ref{fig:Entropy1}
shows that prediction can reduce the entropy by up to 33 bits per
frame. The use of inter-frame prediction results in a reduction of
12 bits compared to coding frames independently. Based on Fig.~\ref{fig:Entropy1},
we select $\alpha=0.8$, for which $\beta=0.6$ is optimal. This provides
close to optimal performance while being more robust to packet loss
than higher values of $\alpha$.

We have found experimentally that the distribution of the prediction
error $q$ is close to a generalised Gaussian distribution of the
form\begin{equation}
N\left(E_{dB}\right)\propto e^{-\left|\frac{E-\mu}{\sigma}\right|^{\gamma}}\ ,\label{eq:sub-Gaussian}\end{equation}
with $\gamma\simeq1.5$. While using~\eqref{eq:sub-Gaussian} directly
in the entropy coder would result in the minimal average bit-rate
for encoding the energy, we prefer to use a Laplace distribution.
By overestimating the least probable values, the Laplace distribution
yields a more constant bit-rate over time, with very little penalty
to the average bit-rate. The achieved rate is only 2~bits above the
lower bound. Exponential-Golomb codes~\cite{Golomb66} could encode
the Laplace-distributed variables with little penalty, but since the
innovation ($Q_{3}$) requires a range coder, we re-use that here.

Once the energy is quantised and encoded at a coarse 6~dB resolution,
a finer scalar quantisation step (with equiprobable symbols) is applied
to achieve a variable resolution that depends on the frequency and
the bit-rate. We use this coarse-fine quantisation process for two
reasons. First, it ensures that most of the information is encoded
in a few bits, which can easily be protected from transmission errors.
Second, it allows us to adjust the fine quantisation of the energy
information to control the total rate allocated to the energy. We
have determined empirically that best results are obtained when the
energy encoding uses about 1/5 of the total allocated bits.%
{}

\subsection{Pitch Gain Quantisation ($Q_{2}$)}

As described in Section~\ref{sub:Pitch-Prediction}, we compute pitch
gains for each band below 8~kHz. While there is a large correlation
between the gains, simple prediction as used for $Q_{1}$ is insufficient.
Instead, the gains are vector-quantised using a 128-entry codebook
(7 bits). To limit the size of the codebook, only 8 values per entry
are considered, so some adjacent bands are forced to have the same
quantised gain. Using 8 bits for each value, the codebook requires
only 1024 bytes of storage. When a special entry in the codebook composed
of all zeros is used, the pitch period does not need to be encoded.
Because the pitch gain is more sensitive to errors when its value
is close to one, we optimise the codebook in the \emph{warped} domain:\begin{equation}
g_{a}^{\left(w\right)}=1-\sqrt{1-g_{a}^{2}}\ .\end{equation}
The codebook is trained and stored in the warped domain $g_{a}^{\left(w\right)}$
so that we can search the codebook using the Euclidean distance metric.
Once the quantised warped gains $\tilde{g}_{a}^{\left(w\right)}$
are found, the quantised gains are \emph{unwarped} by\[
\tilde{g}_{a}=\sqrt{1-\left(1-\tilde{g}_{a}^{\left(w\right)}\right)^{2}}\ .\]

\subsection{Innovation Quantisation ($Q_{3}$)\label{sub:Innovation-Quantisation}}

Because of the normalisation used in CELT, the innovation data lies
on the surface of a hypersphere. While no optimal tessellation is
known for a hypersphere in an arbitrary number of dimensions, a good
approximation is a unit pulse codebook where a code vector $\mathbf{y}$
with $K$ pulses is constructed as\begin{equation}
\mathbf{y}=\sum_{k=1}^{K}s^{(k)}\varepsilon_{n^{(k)}}\ ,\end{equation}
where $n^{(k)}$ and $s^{(k)}$ are the position and sign of the $k^{th}$
pulse, respectively, and $\varepsilon_{n^{(k)}}$ is the $n^{(k)}$th
elementary basis vector. The signs $s_{k}$ are constrained such that
$n^{(j)}=n^{(k)}$ implies $s^{(j)}=s^{(k)}$.

We search the codebook by minimising the square error between the
residual $\mathbf{r}=\mathbf{x}-g_{a}\mathbf{p}$ and $\mathbf{y}$:\begin{align}
\mathbf{y}= & \underset{\mathbf{y}}{\mathrm{argmin}}\left\Vert \mathbf{r}-g_{f}\mathbf{y}\right\Vert ^{2}\ ,\label{eq:energy-min1}\\
= & \underset{\mathbf{y}}{\mathrm{argmin}}\left(\mathbf{r}^{T}\mathbf{r}-2g_{f}\mathbf{r}^{T}\mathbf{y}+g_{f}^{2}\mathbf{y}^{T}\mathbf{y}\right)\ ,\label{eq:energy-min2}\\
= & \underset{\mathbf{y}}{\mathrm{argmin}}\left(\mathbf{r}^{T}\mathbf{r}+J\right)\ ,\\
J= & -2g_{f}\mathbf{r}^{T}\mathbf{y}+g_{f}^{2}\mathbf{y}^{T}\mathbf{y}\ .\label{eq:cost-function}\end{align}
We only need to calculate $J$. The constant term, $\mathbf{r}^{T}\mathbf{r}$,
can be omitted. 

We perform the search one pulse at a time, constraining the sign to
match that of $\mathbf{r}$ at each pulse position. Assuming that
we have already selected $\left(k-1\right)$ pulses, we choose the
next pulse position $n^{\left(k\right)}$ by optimising~\eqref{eq:energy-min2}.
Let $R_{yp}^{\left(k\right)}=\mathbf{p}^{T}\mathbf{y}^{\left(k\right)}$
with $\mathbf{y}^{\left(k\right)}$ containing $k$ pulses and define
$R_{ry}^{(k)}$ and $R_{yy}^{(k)}$ similarly. Then the corresponding
$J^{\left(k\right)}$ can be computed efficiently for each new pulse
using the recursion \begin{align}
s^{(k)}= & \mathrm{sign}\left(r_{n^{(k)}}\right)\ ,\\
R_{yp}^{\left(k\right)}= & R_{yp}^{(k-1)}+s^{(k)}p_{n^{(k)}}\ ,\\
R_{ry}^{\left(k\right)}= & R_{ry}^{(k-1)}+s^{(k)}r_{n^{(k)}}\ ,\\
R_{yy}^{\left(k\right)}= & R_{yy}^{(k-1)}+2s^{(k)}y_{n^{(k)}}^{(k-1)}+1\ ,\\
g_{f}^{\left(k\right)}= & \frac{\sqrt{\tilde{g}_{a}^{2}\left(R_{yp}^{\left(k\right)}\right)^{2}+R_{yy}^{\left(k\right)}\left(1-g_{a}^{2}\right)}-\tilde{g}_{a}R_{yp}^{\left(k\right)}}{R_{yy}^{(k)}}\ ,\\
J^{\left(k\right)}= & -2g_{f}R_{ry}^{\left(k\right)}+g_{f}^{2}R_{yy}^{\left(k\right)}\ .\end{align}

Again, if we have $\tilde{g}_{a}=0$, then the cost function simplifies
to the standard cost function $J=-\mathbf{r}^{T}\mathbf{y}/\sqrt{\mathbf{y}^{T}\mathbf{y}}$.
Failing to take into account the pitch gain and using the standard
cost function yields poorer performance than not using a pitch predictor
at all, since a small error in the fixed codebook contribution may
result in a large final error after~\eqref{eq:fixed-codebook-gain}
is applied if the adaptive codebook contribution is large.

The complexity of the search described above is $\mathcal{O}\left(KN\right)$.
Assuming that the number of bits it takes to encode $K$ pulses is
proportional to $K\log_{2}N$%
\footnote{As we will see in Section \ref{sub:Pulse-Vector-Encoding}, this is
only an approximation.%
}, we can rewrite the complexity of searching a codebook with $b$
bits as $\mathcal{O}\left(bN/\log_{2}N\right)$. By comparison, the
complexity involved in searching a stochastic codebook with $b$ bits
is $\mathcal{O}\left(2^{b}N\right)$, which is significantly higher.

While the structure of the pulse codebook we use has similarities
with ACELP~\cite{Laflamme1990}, the search in CELT is direct and
does not involve filtering operations. On the other hand, the cost
function is more complex, since the fixed codebook gain depends on
both the pitch gain and the code vector selected.

\subsubsection{Reduced search complexity}

For large codebooks, the complexity of the search procedure described
above can be high. We adopt two strategies to reduce that complexity:
\begin{itemize}
\item selecting more than one pulse at a time when $K\gg N$, and
\item using the simpler cost function $J=-\mathbf{r}^{T}\mathbf{y}/\sqrt{\mathbf{y}^{T}\mathbf{y}}$
for all but the last pulse.
\end{itemize}
When the number of pulses is large compared to the number of samples
in a band, we can have a large number of pulses in each position --
in some cases, up to 64 pulses in only 3 positions. Clearly, when
starting the search, there is little risk in assigning more than one
pulse to the position that minimises the cost function in~\eqref{eq:cost-function}.
Therefore in each step we assign\begin{equation}
n_{p}=\max\left(\left\lfloor \left(K-k_{a}\right)/N\right\rfloor ,1\right)\end{equation}
pulses, where $k_{a}$ is the number of pulses that have already been
assigned and $\left\lfloor \cdot\right\rfloor $ denotes truncation
towards zero.

Although using $J=-\mathbf{r}^{T}\mathbf{y}/\sqrt{\mathbf{y}^{T}\mathbf{y}}$
as the cost function reduces quality, it is possible to find most
pulses with it and to use the correct cost function only when placing
the last pulse. This results in a speed gain without any significant
quality degradation.%
{}

\subsubsection{Pulse vector encoding\label{sub:Pulse-Vector-Encoding}}

The pulse vector $\mathbf{y}$ found for each band needs to be encoded
in the bit-stream. We assign a unique index to each possible $\mathbf{y}$
by recursively partitioning the codebook one pulse position at a time~\cite{Fisher1986}.%
{} For $K$ pulses in $N$ samples, the number of codebook entries is\begin{multline}
V\left(N,K\right)=V\left(N-1,K\right)\\
+V\left(N,K-1\right)+V\left(N-1,K-1\right)\ ,\end{multline}
with $V\left(N,0\right)=1$ and $V\left(0,K\right)=0,\ K\neq0$. The
factorial pulse coding (FPC) method~\cite{ACMP00} also achieves
a one-to-one and onto map from pulse vectors to an index less than
$V\left(N,K\right)$, and using FPC would produce bit-identical decoded
output, even though the compressed bit-stream would differ. The main
advantage of the index assignment method used in CELT is that it does
not require multiplications or divisions, and can be implemented without
a lookup table.

The size of a codeword, $\log_{2}V\left(N,K\right)$, is generally
not an integer. To avoid rounding up the the next integer and wasting
an average of half a bit per band (10 bits per frame), we encode the
integers using the range coder. We use equiprobable symbols, so the
encoded size is perfectly predictable. The total overhead of this
method is at most one bit for all bands combined~\cite{Derf2008range},
as has been observed in practice. We have also found that the loss
due to the use of equiprobable symbols (as opposed to using the measured
probability of each symbol) for encoding the innovation is negligible,
as one would expect for a well-tuned vector quantisation codebook.
For the complete codec operating at 64~kbit/s, we have measured that
pulse vector coding results in a saving of 10.8~kbit/s when compared
to encoding each scalar $y_{k}$ value independently using an optimal
entropy coder.

\subsubsection{Sparseness prevention\label{sub:Sparseness-prevention}}

The lack of pitch prediction above 8 kHz and the small number of pulses
used at these frequencies yields a sparsely quantised spectrum, with
few non-zero values. This causes the {}``birdie'' artifacts commonly
found in low bit-rate MPEG-1 Layer 3 (MP3) encodings. To mitigate
this, we use a folded copy of the lower frequency spectrum for the
adaptive code vector, $\mathbf{p}$. We encode a sign bit to allow
the spectrum to be inverted. The principle is similar to~\cite{Makhoul1979},
but is applied in the MDCT domain. The gain $\tilde{g}_{a}$ of this
adaptive codebook is pre-determined, and depends only on the number
of pulses being used and the width of the band. It is given by\begin{equation}
\tilde{g}_{a}=\frac{N}{N+\delta K}\end{equation}
where $\delta=6$ has been found to provide an acceptable compromise
between avoiding birdies and the harshness that can result from spectral
folding.

\subsection{Bit allocation}

Two of the parameter sets transmitted to the decoder are encoded at
variable rate: the energy in each band, which is entropy coded, and
the pitch period, which is not transmitted if the pitch gains are
all zero. To achieve a constant bit-rate without a bit reservoir,
we must adapt the rate of the innovation quantisation. Since CELT
frames are very short, we need to minimise the amount of side information
required to transmit the bit allocation. Hence we do not transmit
this information explicitly, but rather infer it solely from the information
shared between the encoder and the decoder. We first assume that both
the encoder and the decoder know how many 8-bit bytes are used to
encode the frame. This number is either agreed on when establishing
the communication or obtained during the communication, e.g. the decoder
knows the size of any UDP datagram it receives. Given that, both the
encoder and the decoder can implement the same mechanism to determine
the innovation bit allocation.

This mechanism is based solely on the number of bits remaining after
encoding the energy and pitch parameters. A static table determines
the bit-allocation in each band given only the number of bits available
for quantising the innovation. The correspondence between the number
of bits in a band and the number of pulses is given by \eqref{eq:V_NK}.
For a given number of innovation bits, the distribution across the
bands is constant in time. This is equivalent to using a psychoacoustic
masking curve that follows the energy in each band. Because the bands
have a width of one Bark, the result models the masking occuring within
each critical band, but not the masking across critical bands. Using
this technique, no side information is required to transmit the bit
allocation. 

\begin{figure}
\begin{center}\includegraphics[width=0.8\columnwidth]{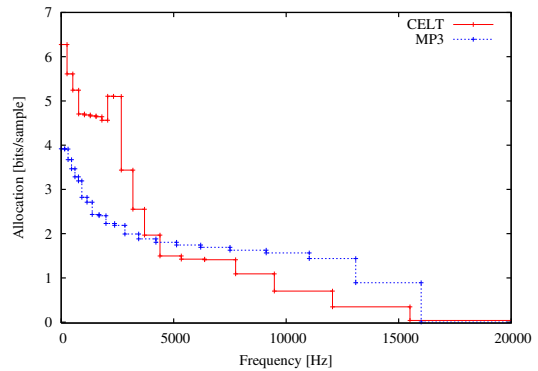}\end{center}

\caption{Average per-band bit allocation for CELT and MP3 at a 64~kbit/s constant
bit-rate. The bit allocation shown is the number of innovation and
fine energy quantisation bits allocated to one band divided by the
number of MDCT bins in that band. The allocation for MP3 includes
the bits from its scale factors that exceed 6dB resolution to ensure
a fair comparison. The piecewise-constant lines also show the division
of the bands for each codec. The average bit-rate for the quantised
MDCT data, excluding the coarse energy/scalefactor quantisation and
any meta-data, is around 55 kbit/s for each codec.\label{fig:bit-distribution}}

\end{figure}

The average bit allocation for all parameters is detailed in Table~\ref{tab:Bit-allocation}.
In addition, Fig. \ref{fig:bit-distribution} shows the average innovation
and fine energy ($Q_{2}$ and $Q_{3}$) bit allocation as a function
of frequency compared to MP3 for a bit-rate of 64~kbit/s. We see
that CELT requires a higher bit-rate to code the low frequencies,
which are often very tonal and difficult to encode with short frames.
On the other hand, the energy constraint allows it to encode the high
frequencies with fewer bits, while still maintaining good quality
(see the next section for a quality comparison). Both codecs use approximately
10~kbit/s for the remaining parameters, despite the fact that MP3
frames are more than double the size of CELT frames (576-sample granules
vs 256-sample frames). %
{}

\begin{table}
\caption{Average bit allocation for each frame. Fractional bits are due to
the high resolution of the entropy coder and to the averaging over
all frames.\label{tab:Bit-allocation}}

\begin{center}\begin{tabular}{ccc}
\hline 
 & \multicolumn{2}{c}{Constant bit-rate}\tabularnewline
\cline{2-3} 
Parameter & 46.9 kbit/s & 64.8 kbit/s\tabularnewline
\hline
Coarse energy ($Q_{1}$) & 36.0 & 35.9\tabularnewline
Fine energy ($Q_{1}$) & 38.1 & 59.2\tabularnewline
Pitch gain ($Q_{2}$) & 7.0 & 7.0\tabularnewline
Pitch period & 8.9 & 8.9\tabularnewline
Innovation ($Q_{3}$) & 180.9 & 264.0\tabularnewline
Unused & 1.1 & 1.0\tabularnewline
\hline
Total per frame & 272 & 376\tabularnewline
\hline
\end{tabular}\end{center}
\end{table}

\section{Evaluation and Discussion}

\label{sec:Experiments-and-Results}

We implemented the CELT codec in C using both floating point and fixed
point. The source code can be obtained at \url{http://www.celt-codec.org/downloads/}
and the results are based on version 0.3.2 of the software. We used
the floating-point version, but the fixed-point version does not cause
noticeable quality degradation. Some audio samples are available at
\url{http://www.celt-codec.org/samples/tasl/} .

\subsection{Other low-delay codecs}

We compare CELT with three other codecs: AAC-LD~\cite{Lutzky2005},
G.722.1C~\cite{Siren14} and MPEG1 Layer III (MP3). The AAC-LD implementation
tested is the one included in Apple's QuickTime Pro (with the {}``best
quality'' option selected). Although AAC-LD has a minimum delay of
20~ms, the Apple implementation uses 512-sample frames and a bit
reservoir, which increases the total delay to 34.8~ms. The G.722.1C
implementation was obtained from the Polycom website%
\footnote{\url{http://www.polycom.com/}%
}. Despite the MP3 codec's high delay, the evaluation included it as
a well known comparison point. We used the LAME MP3 encoder (CBR mode,
with a 20~kHz low-pass filter and no bit reservoir), which significantly
outperformed the dist10 reference MP3 encoder. Table~\ref{tab:Specifications-of-the-codecs}
summarises the characteristics of all the codecs used in the evaluation.
All the codecs compared have at least four times the delay of CELT. 

Unlike AAC-LD and G.722.1C, the Fraunhofer Ultra-Low Delay (ULD) codec~\cite{Wabnik2006}
can achieve coding delays similar to CELT using linear prediction
with pre- and post-filtering \cite{Schuller2002}. Unfortunately,
we were unable to obtain either an implementation or audio samples
for that codec.

\begin{table}
\caption{Characteristics of the codecs at the sampling rate used.\label{tab:Specifications-of-the-codecs} }

\begin{center}\begin{tabular}{ccccc}
\hline 
 & CELT & AAC-LD & G.722.1C & MP3\tabularnewline
\hline
Rate (kHz) & 44.1 & 44.1 & 32 & 44.1\tabularnewline
Frame size (ms) & 5.8 & 10.9 & 20 & variable\tabularnewline
Delay (ms) & 8.7 & 34.8 & 40 & >100\tabularnewline
Bit-rate (kbit/s) & 32-96 & 32-64 & 24,32,48 & 32-160\tabularnewline
\hline
\end{tabular}\end{center}
\end{table}

\subsection{Subjective evaluation}

Untrained listeners evaluated the basic audio quality of the codecs
using the MUlti Stimulus test with Hidden Reference and Anchor (MUSHRA)%
\footnote{Using the RateIt graphical interface available at \url{http://rateit.sf.net/}%
}~\cite{BS1534} methodology. They were presented with audio samples
compressed with CELT, AAC-LD, G.722.1C, and MP3, in addition to low-pass
anchors at 3.5~kHz and 7~kHz. The 7~kHz anchor is the upper bound
achievable by wideband codecs such as G.722, AMR-WB, and G.729.1.

\begin{figure}
\begin{center}\includegraphics[width=1\columnwidth]{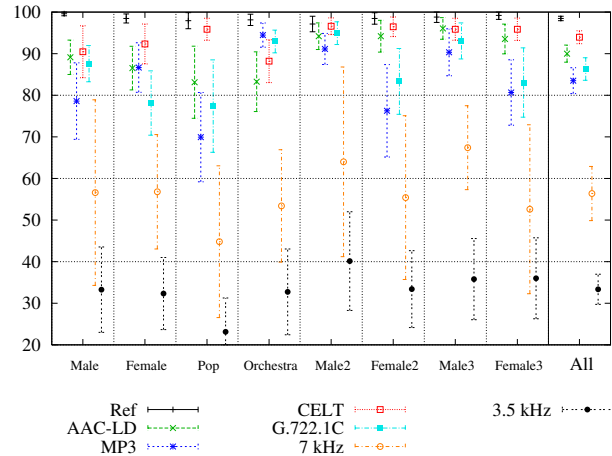}\end{center}

\caption{MUSHRA listening test results at 48~kbit/s with 95\% confidence intervals.\label{fig:MUSHRA-48}}

\end{figure}

The first test included mostly speech samples, divided equally between
male and female and encoded at 48~kbit/s. We used 2 British English
speech samples from the EBU Sound Quality Assessment Material (SQAM)
and 4 American English speech samples from the NTT Multi-Lingual Speech
Database for Telephonometry%
\footnote{We recovered the 44.1 kHz speech from the audio CD tracks and applied
a notch filter to remove an unwanted 15.7 kHz tone from the recording. %
}. The test also included two music samples: a pop music excerpt (Dave
Matthews Band) and an orchestra excerpt (Danse Macabre). For this
test, the CELT codec used 46.9~kbit/s (34 bytes per frame). Fig.~\ref{fig:MUSHRA-48}
shows the average ratings by 15 untrained listeners%
\footnote{Due to an initial problem in generating the 7~kHz anchor, we only
include results for it from the 5 listeners who took the test after
the error was discovered. Results for other codecs and the 3.5 kHz
anchor were not affected by this error and represent data from all
15 listeners.%
}.

\begin{figure}
\begin{center}\includegraphics[width=1\columnwidth]{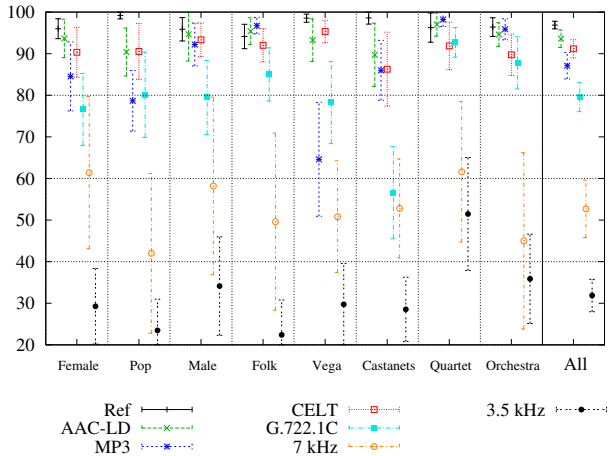}\end{center}

\caption{MUSHRA listening test results at 64~kbit/s with 95\% confidence intervals.\label{fig:MUSHRA-64}}

\end{figure}

A second listening test, at 64~kbit/s, included the following samples:
male speech (SQAM), female speech (SQAM), \emph{a cappella} singing
(Suzanne Vega), vocal quartet (SQAM), pop music (Dave Matthews Band),
folk music (Leahy), castanets (SQAM), and orchestra (Danse Macabre).
For this test, the exact CELT bit-rate was 64.8~kbit/s (47 bytes
per frame). The G.722.1C codec used 48~kbit/s, as this is the highest
bit-rate it supports. Fig.~\ref{fig:MUSHRA-64} shows the average
ratings for these samples were rated by the same 15 untrained listeners
as in the previous test.

The error bars shown in Fig. \ref{fig:MUSHRA-48} and \ref{fig:MUSHRA-64}
represent the 95\% confidence interval for each codec, independently
of the other codecs. However, since listeners were always directly
comparing the same sample encoded with all codecs, a paired statistical
test gives a better assessment of the statistical significance. At
48 kbit/s, CELT is found to out-perform all other codecs with greater
than 95\% confidence when using a paired permutation test (a paired
t-test also shows greater than 95\% confidence). At 64 kbit/s, the
same statistical tests show that CELT out-performs all codecs except
AAC-LD. The results for AAC-LD and G.722.1C are consistent with those
reported at 32~kbit/s in~\cite{Lutzky2005} although in our 64~kbit/s
test, G.722.1C was at a disadvantage, because it was the only codec
operating at 48~kbit/s, the highest rate it supported. 

The 7~kHz low-pass anchors included in the listening tests are equivalent
to an uncompressed signal sampled at 16~kHz, which is the upper bound
achievable by any wideband codec, such as G.722 and AMR-WB. Figs.~\ref{fig:MUSHRA-48}
and~\ref{fig:MUSHRA-64} clearly show that listeners have a very
strong preference for CELT and the other codecs over the 7~kHz low-pass
anchor, demonstrating the benefit of a high sampling rate.

\subsection{Error robustness}

The next experiment measured the robustness of CELT to packet loss
(frame erasure) and bit errors on speech data using the Perceptual
Evaluation of Speech Quality (PESQ)~\cite{P.862} algorithm after
down-sampling the decoder output to 8~kHz. We used PESQ because most
other objective quality evaluation tools, such as PEAQ~\cite{BS1387},
are not designed to estimate quality in noisy channels. The test included
144 files from the NTT multilingual speech database, each from different
speakers (72 male and 72 female), and 18 different languages.

The CELT codec is designed to be robust to packet loss. After a lost
packet, two predictors need to be re-synchronised: the energy predictor
and the pitch predictor. The re-synchronisation time of the energy
predictor depends on the value $\alpha$, while that of the pitch
predictor depends on the pitch gain and period used in subsequent
frames. In practise the re-synchronisation time is limited by the
pitch predictor in voice segments. 

\begin{figure}
\begin{center}\includegraphics[width=0.75\columnwidth]{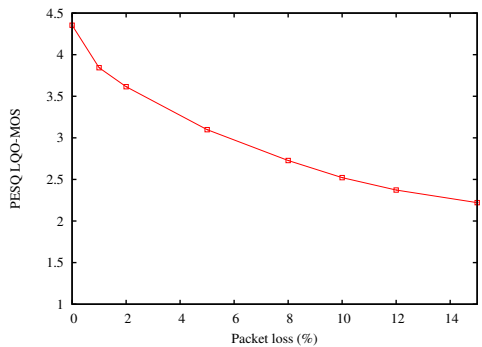}\end{center}\caption{PESQ LQO-MOS as a function of the random packet loss rate. \label{fig:PLC-MOS}}

\end{figure}

\begin{figure}
\begin{center}\includegraphics[width=0.7\columnwidth]{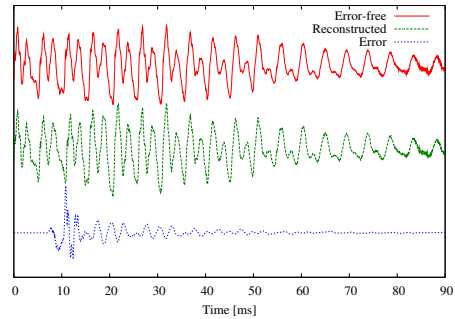}\end{center}\caption{Decoder re-synchronisation after a missing frame. (top) Error-free
decoding (middle) Reconstructed with missing packet (bottom) difference.
\label{fig:Packet-loss} }

\end{figure}

Fig.~\ref{fig:PLC-MOS} shows the PESQ LQO-MOS quality as a function
of the random packet loss rate. The quality remains good at 5\% random
loss and degrades gracefully at higher loss rates. Fig.~\ref{fig:Packet-loss}
shows the recovery from a lost packet during a voice segment. The
recovery time is similar to that obtained by CELP codecs, which are
also limited by pitch prediction. Informal listening tests verified
that most speech utterances remain intelligible up to around 30\%
packet loss. Results for the Fraunhofer ULD codec show quality degradation
on a MUSHRA test with 0.5\% packet loss \cite{Wabnik2005Packet},
the highest level they tested. However, one cannot infer the relative
quality of CELT from this, as MUSHRA and PESQ MOS are not directly
comparable.

\begin{figure}
\begin{center}\includegraphics[width=0.75\columnwidth]{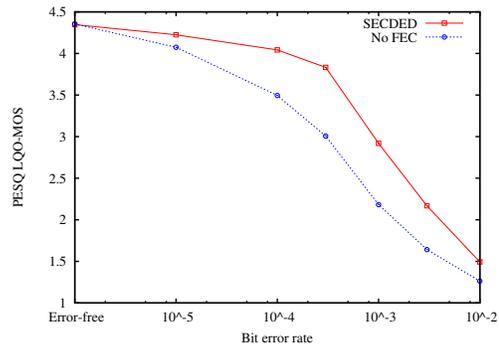}\end{center}\caption{PESQ LQO-MOS as a function of the bit error rate both with SECDED
on the the first 64 bits and without any FEC.\label{fig:PESQ-BER}}

\end{figure}

Although robustness to bit errors is not a specific aim for the codec,
we performed limited testing in bit-error conditions at 46.9~kbit/s
(34~bytes per frame). Testing was performed in two different conditions:
without any forward error correction (FEC) and with a simple 8-bit
single-error-correcting, double-error-detecting (SECDED) code applied
to the first 64~bits, which mainly consist of energy and pitch information.
In cases where a double error was detected, the frame was considered
lost. An evaluation of the speech quality as a function of the BER
in Fig. \ref{fig:PESQ-BER} shows that robustness up to a $3\times10^{-4}$
BER can be achieved at the cost of an 8~bit per frame (1.4~kbit/s)
reduction in the codec's base bit-rate. Since the CELT bit-rate can
be adjusted dynamically, a good strategy for transmission over a noisy
channel, e.g. a wireless link, would be to adapt the bit-rate to the
channel capacity, as the AMR codecs do~\cite{Bessette2002}.

\subsection{Complexity}

The floating point version of the codec requires approximately 30~MFLOPS
for encoding and decoding in real-time at 44.1~kHz, or around 5\%
of a single core on a 2~GHz Intel Core~2 CPU, without CPU-specific
optimisation. When implemented in fixed-point on a Texas Instruments
TMS320C55x-family DSP, it requires 104~MIPS to perform both encoding
and decoding in real-time, using 7.7~kWords (15.4~kB) of data RAM.
Table~\ref{tab:Implementation-complexity} gives more details of
the complexity. This makes CELT comparable to AAC-LD, although more
complex than G.722.1C, which has very low complexity.

\begin{table}
\caption{Implementation complexity of CELT on a TI TMS320C55x-family DSP. Per-channel
data is persistent from one frame to another, while scratch data is
only required while the codec is executing.\label{tab:Implementation-complexity}}

\begin{center}\begin{tabular}{lccc}
\hline 
 & Encoder & Decoder & Both\tabularnewline
\hline
Computation (MIPS) & 68 & 36 & 104\tabularnewline
Per-channel RAM (kB) & 5.1 & 4.6 & 9.7\tabularnewline
Scratch data RAM (kB) & 5.7 & 2.6 & 5.7\tabularnewline
Table data ROM (kB) & 6 & 3 & 6\tabularnewline
\hline
\end{tabular}\end{center}
\end{table}

\subsection{Reducing the delay}

\begin{figure}
\begin{center}\includegraphics[width=0.75\columnwidth]{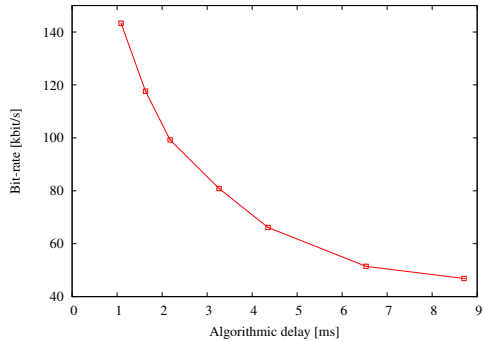}\end{center}

\caption{Bit-rate required as a function of the algorithmic delay to achieve
a constant quality.\label{fig:Bit-rate-vs-delay}}

\end{figure}

We have shown results here for CELT operating with 256-sample frames
and a 384-sample total algorithmic delay (8.7~ms at 44.1~kHz). The
same codec can be used with even smaller frame sizes. Fig.~\ref{fig:Bit-rate-vs-delay}
shows the bit-rate required to achieve a constant quality level when
lowering the algorithmic delay. The reference quality is that obtained
at 46.9~kbit/s with 8.7~ms delay and is measured using PQevalAudio%
\footnote{\url{http://www-mmsp.ece.mcgill.ca/Documents/Software/Packages/AFsp/PQevalAudio.html}%
}, an implementation of the PEAQ basic model~\cite{BS1387}. We observe
that CELT scales well down to 3~ms delay, at which point the required
bit-rate goes up very quickly. This is largely because the cost of
encoding the band energies and the pitch information is nearly constant
per frame, regardless of the frame size.

\section{Conclusion}

\label{sec:Discussion-and-Conclusion}

We proposed a new constrained-energy lapped transform (CELT) structure
for speech coding at high sampling rates and very low-delay. The CELT
algorithm can achieve high-quality coding at low delay by using an
efficient algebraic shape-gain quantiser that preserves the spectral
envelope of the signal, while minimising the side information transmitted.
Additionally, a time-domain pitch predictor partially compensates
for the poor frequency resolution obtained with the short MDCT windows.
Results show that at 48~kbit/s and 64~kbit/s, CELT out-performs
G.722.1C and MP3 on our test data and is comparable to AAC-LD, despite
having less than one fourth of the algorithmic delay of the codecs
to which it was compared.

There are still several ways to improve CELT, such as by incorporating
better psychoacoustic masking in the dynamic bit allocation. This
is a difficult problem both because there are few bits available for
coding the allocation and because the analysis window is short. 

\appendices

\section{Example bit-stream}

\begin{table}
\caption{Detailed innovation bit allocation for a frame encoded at 64.8~kbit/s.
For each band, we give the frequency (in MDCT bins) where the band
starts, the width of the band (in MDCT bins), the number of pulses
allocated to the band, and the number of bits required ($\log_{2}V\left(N,K\right)$).
Although no pulse is assigned to its innovation, band 19 still uses
one bit for the folding sign (Section \ref{sub:Sparseness-prevention}).
Band 20, corresponding to frequencies above 20~kHz, is not coded
and is set to zero at the decoder.\label{tab:Detailed-innovation-bit}}

\begin{center}\begin{tabular}{ccccc}
\hline 
\multicolumn{5}{c}{Frame \#270 -- Dave Matthews Band}\tabularnewline
\hline
Band & Start & Width ($N$) & Pulses ($K$) & Bits\tabularnewline
\hline
0 & 0 & 3 & 38 & 12.5\tabularnewline
1 & 3 & 3 & 28 & 11.6\tabularnewline
2 & 6 & 3 & 20 & 10.6\tabularnewline
3 & 9 & 3 & 15 & 9.8\tabularnewline
4 & 12 & 3 & 15 & 9.8\tabularnewline
\hline 
5 & 15 & 3 & 15 & 9.8\tabularnewline
6 & 18 & 3 & 15 & 9.8\tabularnewline
7 & 21 & 3 & 14 & 9.6\tabularnewline
8 & 24 & 3 & 20 & 10.6\tabularnewline
9 & 27 & 4 & 28 & 15.8\tabularnewline
\hline 
10 & 31 & 6 & 13 & 17.7\tabularnewline
11 & 37 & 6 & 7 & 13.4\tabularnewline
12 & 43 & 8 & 6 & 14.7\tabularnewline
13 & 51 & 11 & 5 & 15.4\tabularnewline
14 & 62 & 12 & 5 & 16.1\tabularnewline
\hline 
15 & 74 & 16 & 6 & 21.6\tabularnewline
16 & 90 & 20 & 5 & 20.7\tabularnewline
17 & 110 & 30 & 4 & 20.0\tabularnewline
18 & 140 & 40 & 2 & 12.6\tabularnewline
19 & 180 & 53 & 0 & 1\tabularnewline
\hline
(20) & 233 & 23 & not coded & 0\tabularnewline
\hline
\multicolumn{4}{c}{\emph{Innovation total}} & 263.4\tabularnewline
\hline
\multicolumn{4}{c}{\emph{Overhead and padding}} & 1.3\tabularnewline
\hline
\end{tabular}\end{center}
\end{table}

Consider a single (typical) frame from the Dave Matthews Band excerpt
encoded at 64.8~kbit/s. In that frame, the pitch gain is first encoded
using 7 bits. Since the gain is non-zero, the pitch period is then
encoded using 9.3 bits%
\footnote{Fractional bits are possible because we use a range coder with equiprobable
symbols.%
} ($\log_{2}641$). The energy in each band is encoded with 94.7 bits,
followed by the innovation, which requires 263.4 bits, as detailed
in Table \ref{tab:Detailed-innovation-bit}. The innovation bits in
Table \ref{tab:Detailed-innovation-bit} include the sparseness prevention
signs (one each for bands 15-19), as explained in Section \ref{sub:Sparseness-prevention}.
One bit is left unused to account for overhead due to finite precision
arithmetic in the range coder, for a total of 376 bits (47 bytes).
The bit allocation for all other frames is very similar, with the
main variation occuring in frames that do not have a pitch period
(gain is zero).

\bibliographystyle{IEEEbib}
\bibliography{celt}

\begin{biography}[{\includegraphics[width=1in]{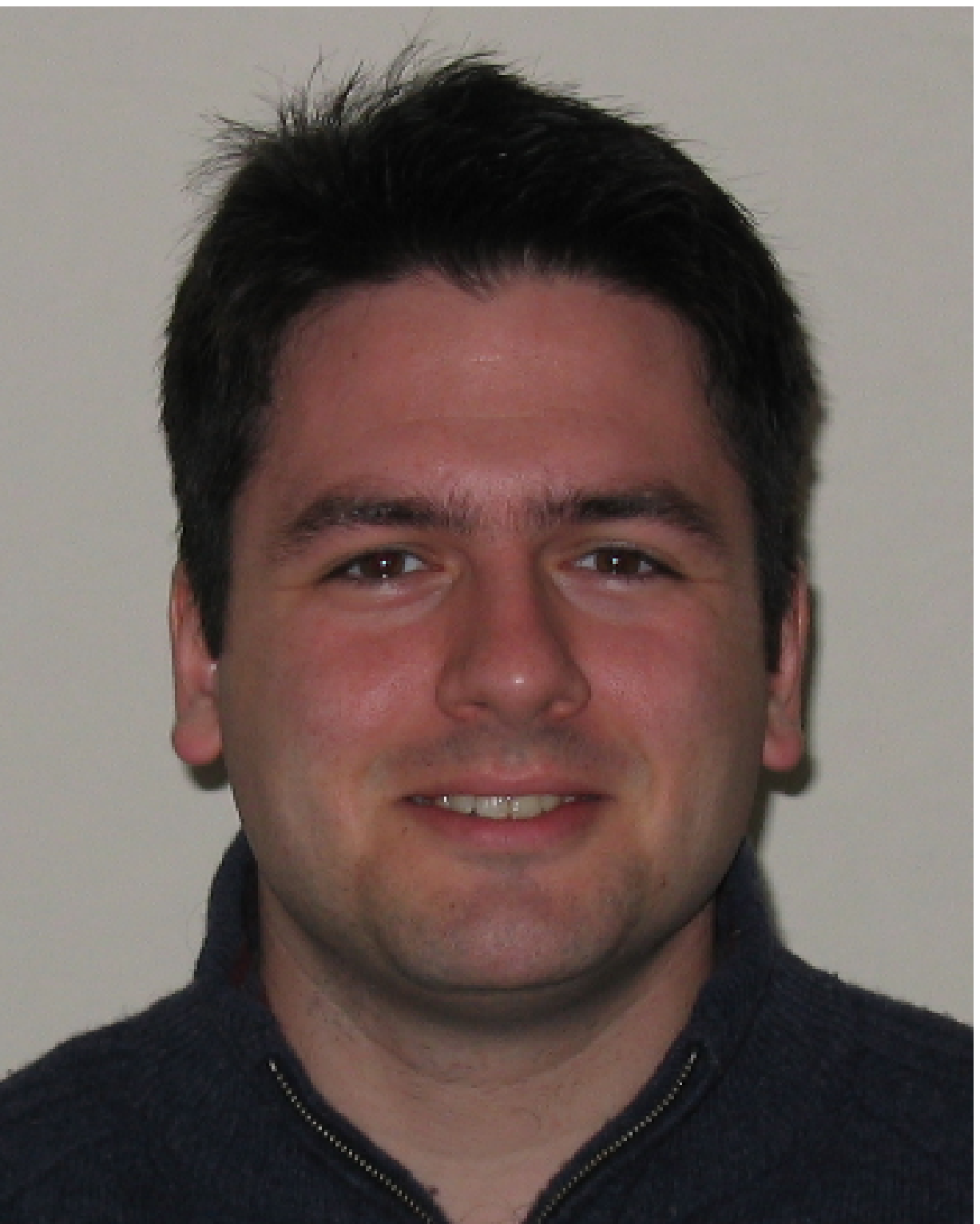}}]{Jean-Marc
Valin} (S'03-M'05) holds B.Eng. ('99), M.A.Sc. ('01) and Ph.D. ('05)
degrees in electrical engineering from the University of Sherbrooke.
His Ph.D. research focused on bringing auditory capabilities to a
mobile robotics platform, including sound source localisation and
separation. In 2002, he authored the Speex open source speech codec,
which he keeps maintaining to this date. Since 2005, he is a software
lead architect at Octasic Inc. and his interests include acoustic
echo cancellation and audio coding.\end{biography}

\begin{biography}[{\includegraphics[width=1in]{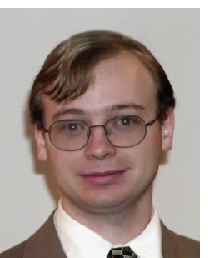}}]{Timothy
B. Terriberry} received dual B.S. and M.S. degrees from Virginia
Tech in 1999 and 2001, respectively, in both Mathematics and Computer
Science, and a Ph.D. in Computer Science from the Univeristy of North
Carolina at Chapel Hill in 2006. He has volunteered for the Xiph.Org
Foundation -- a non-profit organization that develops free, open multimedia
protocols and software -- since 2002.\end{biography}

\begin{biography}[{\includegraphics[width=1in]{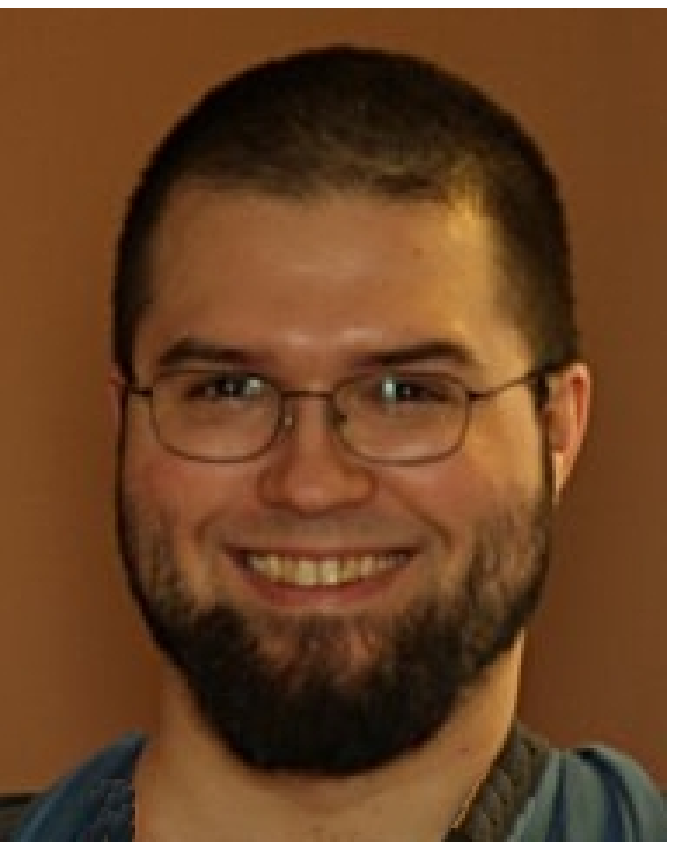}}]{Christopher
(Monty) Montgomery} founded the Xiph.Org Foundation and authored
Ogg Vorbis and other open-source packages. He holds a B.S. in Electrical
Engineering and Computer Science from MIT and a M.Eng. in Computer
Engineering from the Tokodai in Japan. He is currently a Senior Engineer
at Red Hat. \end{biography}

\begin{biography}[{\includegraphics[width=1in]{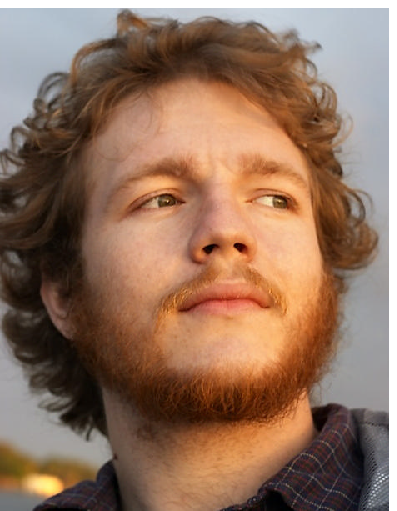}}]{Gregory
Maxwell} is a Senior Systems Engineer for Juniper Networks in Herndon,
Virginia and has volunteered for the Xiph.Org Foundation since 1999.
His interests include spatial audio, audio compression, and radio
systems. \end{biography}

\vfill{}

\end{document}